\documentclass{elsart}

\usepackage{graphicx,natbib,amssymb,lineno,hyperref}
\usepackage{lscape}

\def\astrobj#1{#1}

\newcommand\apj{{ApJ}}%
\newcommand\aj{{AJ}}%
\begin{document}

\begin{frontmatter}
\title{The data mining III: An analysis of 21 eclipsing binary light-curves observed by the INTEGRAL/OMC}

\author{P. Zasche}
\ead{zasche@sirrah.troja.mff.cuni.cz}

\address{Astronomical Institute, Faculty of Mathematics and Physics,
 Charles University Prague, CZ-180 00 Praha 8, V Hole\v{s}ovi\v{c}k\'ach 2, Czech Republic}

\begin{abstract}
Twenty-one eclipsing binaries were selected for an analysis from a huge database of observations
made by the INTEGRAL/OMC camera. The photometric data were processed and analyzed, resulting in a
first light-curve study of these neglected eclipsing binaries. In several systems from this sample
even their orbital periods have been confirmed or modified. 32 new minima times of these binaries
have been derived.
\end{abstract}

\begin{keyword}
stars: binaries: eclipsing \sep stars: individual: \astrobj{V479 Aql}, \astrobj{DY Aqr},
\astrobj{V646 Cas}, \astrobj{PV Cen}, \astrobj{EP Cen}, \astrobj{RW Cet}, \astrobj{RR Cir},
\astrobj{UX Cir}, \astrobj{BB Cir}, \astrobj{UW Cyg}, \astrobj{V445 Cyg}, \astrobj{SX Gem},
\astrobj{AB Mus}, \astrobj{UZ Nor}, \astrobj{XY Nor}, \astrobj{V2383 Oph}, \astrobj{FL Ori},
\astrobj{SW Pup}, \astrobj{V501 Sgr}, \astrobj{V1133Sgr}, \astrobj{EL Vel} \sep stars: fundamental
parameters \PACS 97.10.-q \sep 97.80.-d \sep 97.80.Hn
\end{keyword}

\end{frontmatter}

\section{Introduction}

The photometric observations are obtained by the INTEGRAL (INTErnational Gamma-Ray Astrophysics
Laboratory) satellite since its launch in 2002. The onboard OMC (Optical Monitoring Camera) was
designed to carry out the observations in optical $V$ passband. These observations are in fact only
a by-product of the mission, but nowadays there are many observations available.

This investigation is directly following our previous papers (\citealt{Zasche2008NewA},
\citealt{2009NewA...14..129Z}, and \citealt{2010NewA...15..150Z}). The selection criteria used here
were also the same: maximum number of data points and non-existence of any detailed light-curve
analysis of the particular system. There were selected 21 eclipsing binary systems for the present
paper.

\section{Analysis of the individual systems}

All observations of these systems were carried out by the same instrument (50mm OMC telescope) and
the same filter (standard Johnson's V filter). Time span of the observations ranges from December
2002 to January 2009. A transformation of the time scale has been done following the equation
$Julian Date - ISDC Julian Date = 2,451,544.5$. Only a few outliers from each data set were
excluded. The {\sc Phoebe} programme (see e.g. \citealt{Prsa2005}), based on the Wilson-Devinney
algorithm \citep{Wilson1971}, was used for the analysis.

Due to missing information about the stars, and having only the light curves in one filter, some of
the parameters have to be fixed. At first, for all the systems we have used the "Detached binary"
mode (in Wilson \& Devinney mode 2) and also the "Semidetached with the secondary component filling
its Roche lobe" (mode 5 in Wilson \& Devinney) for computing. For both modes a "q-search method"
was used, which means trying to find the best fit with different values of the mass ratio $q$
ranging from 0 to 1 with a step 0.1. The limb-darkening coefficients were interpolated from
van~Hamme's tables (see \citealt{vanHamme1993}), the linear cosine law was used. The values of the
gravity brightening and bolometric albedo coefficients were set at their suggested values for
convective atmospheres (see \citealt{Lucy1968}), i.e. $G_1 = G_2 = 0.32$, $A_1 = A_2 = 0.5$. In all
cases the orbital eccentricity was set to 0 (circular orbit). Therefore, the quantities which could
be directly calculated from the light curve are the following: The relative luminosities $L_i$, the
temperature ratio $T_1/T_2$, the inclination $i$, ephemerides of the system, the Kopal's modified
potentials $\Omega_1$ and $\Omega_2$, the synchronicity parameters $F_1$ and $F_2$, the third light
$L_3$, and the mass ratio $q$. Using the parameters introduced above, one could also derive the
value of the relative radii $R_i/a$.

%\begin{landscape}
\begin{table*}[t]
 \scriptsize
 \caption{Basic information about the analyzed systems, taken from the literature.}
 \label{Table1} \centering \scalebox{0.74}{
\begin{tabular}{c c c c c c c c c c c c c c}
\hline \hline
    Star   & Mag B & Mag V & (B-V) &  (B-V) & Sp. &    Sp.     &  q   & Type  & Min  &  Mag  &  Mag  & Mag   &\multicolumn{1}{c}{Data}\\[-1mm]
           & GCVS  & GCVS  & GCVS  &  Nomad &     &   S\&K     & S\&K & S\&K  &      &  OMC  &  MinI & MinII &  \\
    \hline
  V479 Aql & 13.20 & 13.2  & 0.0  &  0.28 &      & (A8)+[G9IV] & 0.300 & EA/SD &  76 & 12.99 & 14.14 & 13.10 & 408  \\[-1mm]
    DY Aqr & 10.30 & 10.23 & 0.07 &  0.14 &  A0  &   A0+[F3]   & 0.540 & EA/DM &  35 & 10.22 & 10.87 & 10.26 & 479  \\[-1mm]
  V646 Cas & 10.14 & 9.73  & 0.58 &  0.46 & B0IV &             &       &       &  24 &  9.53 &  9.97 &  9.76 & 756  \\[-1mm]
    PV Cen & 11.40 & 11.4  & 0.0  &  0.21 &      & (A2)+[G4IV] & 0.270 & EA/SD &   2 & 11.55 & 12.25 & 12.05 & 623  \\[-1mm]
    EP Cen & 11.30 & 11.3  & 0.0  & -0.4  &      & (F0)+[G4IV] & 0.590 & EA/SD &   3 & 11.18 & 11.69 & 11.27 & 470  \\[-1mm]
    RW Cet & 10.43 & 10.09 & 0.34 &  0.37 &  A5  &   A5+[G6IV] & 0.400 & EA/SD &  79 & 10.16 & 10.99 & 10.24 & 449  \\[-1mm]
    RR Cir & 11.2  & 10.9  & 0.3  &  0.03 &      &             &       &       &   4 & 11.51 & 12.57 & 11.84 & 450  \\[-1mm]
    UX Cir & 12.20 & 12.2  & 0.0  &  0.12 &      &             &       &       &   2 & 11.69 & 12.25 & 11.77 & 455  \\[-1mm]
    BB Cir &  9.40 &  9.4  & 0.0  &  0.38 & A2IV & A2IV+[G9IV] & 0.290 & EA/SD &  32 & 10.04 & 11.55 & 10.11 & 514  \\[-1mm]
    UW Cyg & 11.00 & 10.68 & 0.32 &  0.22 &  A5  &   F0+[K4IV] & 0.280 & EA/SD & 188 & 10.78 & 13.48 & 10.87 & 457  \\[-1mm]
  V445 Cyg & 12.40 & 11.7  & 0.7  &  0.44 &  M5V & (A2)+[G9IV] & 0.200 & EA/SD &  74 & 12.04 & 15.01 & 12.15 & 462  \\[-1mm]
    SX Gem & 11.2  & 11.0  & 0.2  & -0.23 &  A0  &   A0+[A7]   & 0.710 & EA/DM & 110 & 11.40 & 12.27 & 11.46 & 315  \\[-1mm]
    AB Mus & 12.80 &       &      &  0.47 &      & (A8)+[G3IV] & 0.520 & EA/SD &   1 & 13.38 & 14.58 & 13.50 & 469  \\[-1mm]
    UZ Nor & 11.20 & 11.2  & 0.0  &  0.06 &      & (A1)+[K0IV] & 0.180 & EA/SD &   0 & 11.04 & 14.42 & 11.13 & 467  \\[-1mm]
    XY Nor & 13.00 & 13.0  & 0.0  &  0.02 &  A6  & (A5)+[F1.5] & 0.720 & EA/DM &  16 & 12.47 & 12.94 & 12.72 & 502  \\[-1mm]
 V2383 Oph & 11.5  & 10.3  & 1.2  &  1.21 &  K7V &             &       &       &   8 & 10.29 & 10.98 & 10.57 & 886  \\[-1mm]
    FL Ori & 11.5  & 11.5  & 0.0  &  0.32 &  A3V &  A3V+[K0IV] & 0.200 & EA/SD & 185 & 11.05 & 12.10 & 11.14 & 472  \\[-1mm]
    SW Pup &  9.30 & 9.01  & 0.29 &  0.30 &  A0V &  A0V+[F8IV] & 0.460 & EA/SD &   2 &  8.97 &  9.96 &  9.08 & 380  \\[-1mm]
  V501 Sgr & 12.60 & 12.6  & 0.0  & -0.22 &      & (A3)+[G8IV] & 0.230 & EA/SD &   0 & 12.36 & 13.07 & 12.40 & 469  \\[-1mm]
  V1133Sgr & 13.20 & 13.2  & 0.0  &  0.32 &      & (F0)+[G8IV] & 0.300 & EA/SD &   2 & 12.49 & 13.47 & 12.57 & 317  \\[-1mm]
    EL Vel & 11.7  & 11.3  & 0.4  &  0.30 &      & (A3)+[G5IV] & 0.150 & EA/SD &   2 & 11.32 & 12.34 & 11.36 & 422  \\ \hline
\end{tabular}}
\end{table*}
%\end{landscape}

The distinguishing between the primary and secondary minima has been done only according to the
observational point of view, which means that the deeper one is the primary one. This results in a
fact that the primary component could be neither the larger one, nor the more massive one. In one
case the secondary component resulted to be the more luminous one (V646~Cas), and in V1133~Sgr the
more massive one.

The basic information about the analyzed systems are introduced in Table \ref{Table1}, where are
the $B$ and $V$ magnitudes from the GCVS (\citealt{1971GCVS} and \citealt{Malkov}), the $B-V$
values from the GCVS and also from the NOMAD catalogue \citep{NOMAD2004}. The spectral types are
taken from the published literature and also from the \citeauthor{Svechnikov} (S\&K,
\citealt{Svechnikov}). The estimated mass ratio and also the type of the eclipsing binary have been
taken from S\&K (EA stands for the Algol type, SD for semi-detached systems, and DM for detached
main sequence ones). 'Min' stands for the number of published times of minima and the last four
columns introduce the actual OMC magnitudes in Johnson's $V$ filter, the depths of both primary and
also secondary minima in $V$ filter, and finally the number of data points used for this analysis.

\begin{figure}[b]
 \includegraphics[width=16cm]{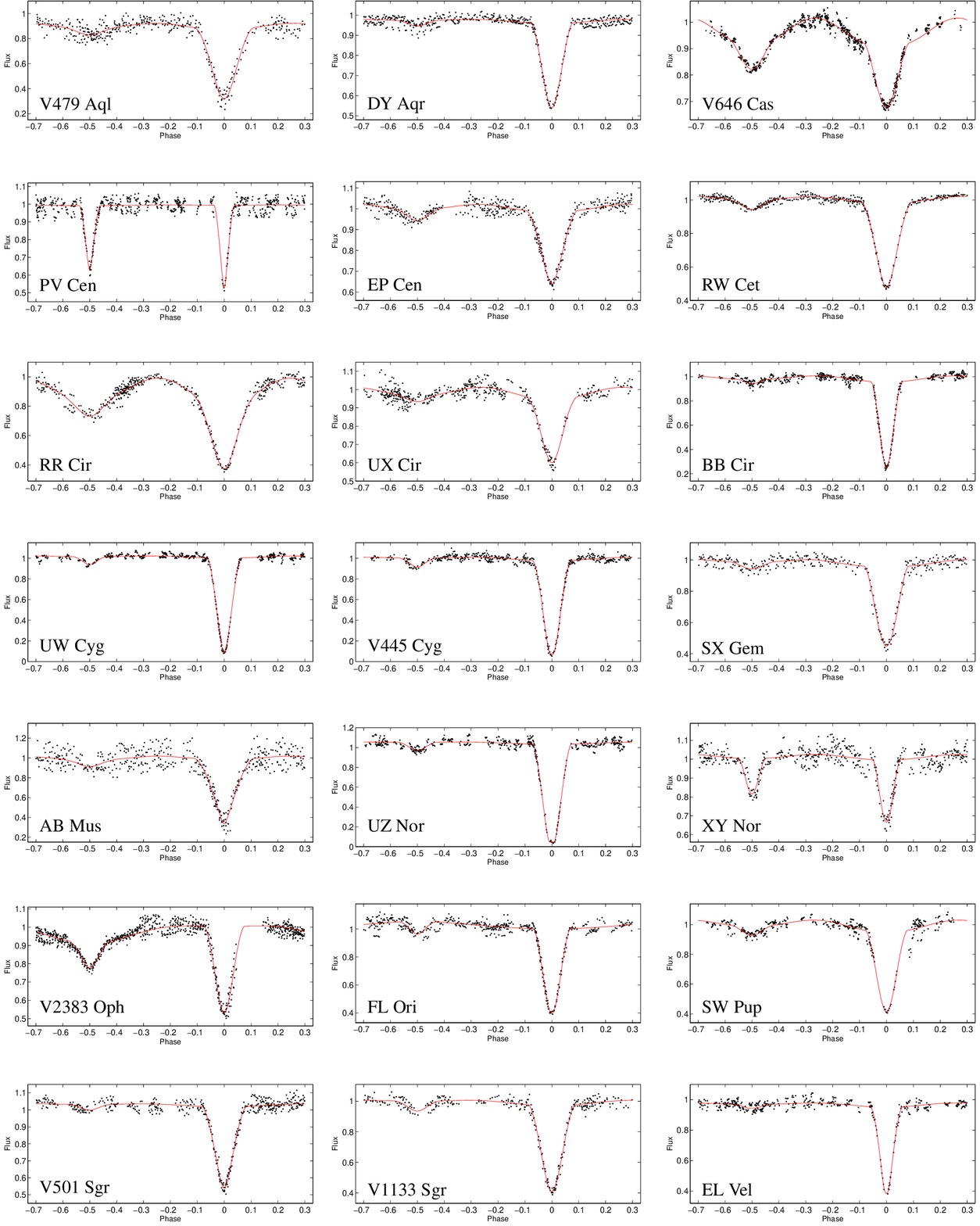}
 \caption{The light curves of the analyzed systems.}
 \label{Figs}
\end{figure}

The results are introduced in Fig.\ref{Figs} and Table \ref{Table2}, where are given all relevant
parameters of the analyzed systems: HJD$_0$ and $P$ are the ephemerides of the system, $i$ stands
for the inclination, $q$ ($=M_2/M_1$) denoted the mass ratio, the 'Type' refers the mode used for
the best solution ('D' for a detached na 'SD' for a semi-detached one, see above), $\Omega_i$
stands for the Kopal's modified potentials, $T_i$ for the effective temperatures, $L_i$ for the
luminosities, $R_i/a$ for the relative radii, $F_i$ for the synchronicity parameters, and $x_i$ for
the limb-darkening coefficients (the linear cosine law was used), respectively. Inclinations
smaller than $90^\circ$ mean that the binary rotates counter-clockwise as projected onto a plane of
sky. In some systems their orbital periods were found to be different from the values published in
the literature (e.g. in GCVS).

%\begin{landscape}
\begin{table*}[t]
 \scriptsize
 \caption{The light-curve parameters of the individual systems, as derived from our analysis.}
 \label{Table2} \centering \scalebox{0.62}{
\begin{tabular}{ c c c c c c c c c r r r c c c c c c }
\hline \hline
 Parameter &  HJD$_0$ &      P     &   $i$  & $q$ &Type&$\Omega_1$& $\Omega_2$ & $T_1/T_2$ &\multicolumn{1}{c}{$L_1$}&\multicolumn{1}{c}{$L_2$}&\multicolumn{1}{c}{$L_3$}& $R_1/a$ & $R_2/a$ & $F_1$ & $F_2$ & $x_1$ & $x_2$ \\[-1mm]
    Star   & 2450000+ &   [days]   &  [deg] &     &    &          &            &           &\multicolumn{1}{c}{[\%]}&\multicolumn{1}{c}{[\%]}&\multicolumn{1}{c}{[\%]}&          &      &       &       &       &       \\
    \hline
  V479 Aql & 2765.422 & 0.83359673 & 80.152 & 0.8 & SD &  4.1135  &     --     &   1.558   & 89.05 &  8.70 &  2.25 & 3.02 & 3.36 & 0.000 & 1.505 & 0.433 & 0.565 \\[-1mm]
    DY Aqr & 2634.187 & 2.15969977 & 80.375 & 0.4 & SD &  4.8440  &     --     &   2.145   & 72.17 &  4.49 & 23.34 & 2.28 & 2.62 & 1.486 & 2.023 & 0.319 & 0.533 \\[-1mm]
  V646 Cas & 3275.126 & 6.16261291 & 69.273 & 0.7 & SD &  5.3695  &     --     &   1.477   & 48.45 & 51.55 &  0.00 & 2.28 & 3.59 & 2.968 & 0.727 & 0.317 & 0.366 \\[-1mm]
    PV Cen & 2837.827 & 3.83468075 & 88.804 & 1.0 & D  &  9.0974  &   9.7506   &   0.932   & 54.77 & 37.80 &  7.43 & 1.24 & 1.14 & 0.166 & 1.125 & 0.681 & 0.623 \\[-1mm]
    EP Cen & 2832.788 & 0.92544905 & 82.534 & 0.5 & SD &  5.4393  &     --     &   1.777   & 40.18 & 10.22 & 49.60 & 2.31 & 3.22 & 4.509 & 0.959 & 0.364 & 0.524 \\[-1mm]
    RW Cet & 3925.933 & 0.97519630 & 82.251 & 0.4 & SD &  4.8092  &     --     &   1.860   & 63.08 &  8.24 & 28.68 & 2.48 & 3.11 & 3.494 & 0.740 & 0.398 & 0.627 \\[-1mm]
    RR Cir & 2834.296 & 1.09172354 & 79.075 & 0.6 & SD &  4.0259  &     --     &   1.197   & 80.88 & 19.12 &  0.00 & 3.33 & 3.57 & 2.474 & 0.020 & 0.771 & 0.500 \\[-1mm]
    UX Cir & 3372.779 & 1.48126165 & 79.854 & 0.7 & SD &  5.8007  &     --     &   1.657   & 55.04 &  7.20 & 37.76 & 2.33 & 3.20 & 4.686 & 1.620 & 0.543 & 0.786 \\[-1mm]
    BB Cir & 3211.257 & 3.08719168 & 83.402 & 0.9 & D  &  6.3804  &   6.3659   &   1.721   & 98.74 &  1.26 &  0.00 & 1.93 & 1.71 & 3.510 & 1.828 & 0.500 & 0.500 \\[-1mm]
    UW Cyg & 2631.030 & 3.45078022 & 85.287 & 0.7 & D  &  6.3882  &   4.7967   &   1.710   & 92.39 &  7.61 &  0.00 & 1.85 & 2.01 & 3.765 & 2.564 & 0.436 & 0.636 \\[-1mm]
  V445 Cyg & 3189.713 & 1.94778095 & 87.306 & 0.7 & D  &  6.7239  &   4.4609   &   1.880   & 87.63 & 12.37 &  0.00 & 1.78 & 2.40 & 4.569 & 3.249 & 0.544 & 0.762 \\[-1mm]
    SX Gem & 3111.556 & 1.36718292 & 76.369 & 0.8 & SD &  6.8650  &     --     &   2.270   & 62.95 & 18.47 & 18.58 & 1.73 & 3.60 & 3.956 & 1.019 & 0.265 & 0.483 \\[-1mm]
    AB Mus & 2834.149 & 0.96494625 & 87.661 & 0.7 & SD &  4.1649  &     --     &   1.042   & 71.31 &  8.59 & 20.10 & 3.02 & 3.16 & 1.658 & 1.718 & 0.142 & 0.148 \\[-1mm]
    UZ Nor & 3059.574 & 3.19609307 & 86.554 & 0.6 & D  &  6.2150  &   3.4925   &   2.112   & 87.66 & 12.34 &  0.00 & 1.78 & 2.65 & 0.000 & 0.794 & 0.266 & 0.395 \\[-1mm]
    XY Nor & 3223.125 & 1.68001447 & 81.806 & 0.9 & D  &  8.1800  &   6.5509   &   1.206   & 53.45 & 46.55 &  0.00 & 1.51 & 1.84 & 6.530 & 5.068 & 0.555 & 0.702 \\[-1mm]
 V2383 Oph & 2911.269 & 0.50220468 & 82.407 & 0.9 & D  &  5.4312  &   6.0771   &   0.683   & 77.40 & 22.60 &  0.00 & 2.39 & 2.12 & 2.974 & 4.792 & 0.500 & 0.500 \\[-1mm]
    FL Ori & 4147.705 & 1.55098194 & 84.512 & 0.9 & D  &  6.7050  &   4.8162   &   2.349   & 63.44 & 11.99 & 24.57 & 1.73 & 2.51 & 0.947 & 2.049 & 0.337 & 0.649 \\[-1mm]
    SW Pup & 2979.169 & 2.74742494 & 76.517 & 0.8 & D  &  7.2531  &   3.7446   &   1.774   & 80.99 & 19.01 &  0.00 & 1.86 & 3.15 & 6.559 & 1.234 & 0.409 & 0.608 \\[-1mm]
  V501 Sgr & 3447.115 & 1.51327251 & 86.812 & 0.8 & D  &  5.7024  &   4.3304   &   2.345   & 49.40 &  3.21 & 47.39 & 2.19 & 2.66 & 3.266 & 2.187 & 0.276 & 0.522 \\[-1mm]
  V1133Sgr & 4377.904 & 0.80532386 & 81.079 & 1.1 & D  &  6.3172  &   4.3754   &   1.871   & 63.05 & 11.71 & 25.24 & 2.09 & 3.21 & 3.573 & 0.000 & 0.398 & 0.635 \\[-1mm]
    EL Vel & 2974.883 & 2.75833767 & 91.351 & 0.3 & D  &  6.0711  &   3.4474   &   2.162   & 96.00 &  0.86 &  3.14 & 2.03 & 1.48 & 6.360 & 2.654 & 0.771 & 0.500 \\ \hline
\end{tabular}}
\end{table*}
%\end{landscape}

\begin{figure}
 \includegraphics[width=12cm]{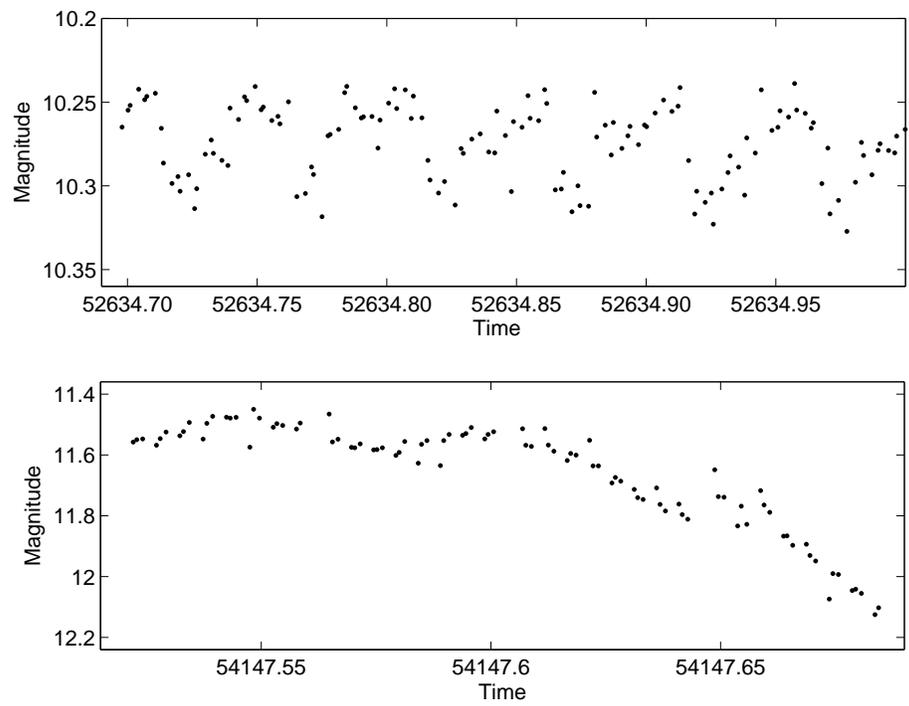}
 \caption{The observed pulsations in DY Aqr (top) and FL Ori (bottom).}
 \label{Figs2}
\end{figure}

In two systems (DY Aqr and FL Ori) there were found short-periodic pulsations. In Fig. \ref{Figs2}
are presented parts of the light curve with evident pulsational behavior. Pulsations in FL Ori have
been predicted by \cite{2006MNRAS.370.2013S}. On the other hand, the proposed light variations in
RW~Cet have not been detected in the present data.

Another interesting fact of this sample is that about one half of the investigated systems have the
luminosity of the third unseen body above a statistically significant value about 5\%. This result
is not surprising, because e.g. \cite{Pribulla} also discovered that more than 50\% of binaries
exist in multiple systems. One could speculate about a prospective future discovery of such
components in these systems. Due to missing detailed analysis (spectroscopic, interferometric,
etc.), the only possible way how to discover these bodies nowadays is the period analysis of their
times of minima variations. In the systems RW~Cet and FL~Ori there exist some indication of period
modulation during the last decades. New minima times for some of the systems have been derived from
the OMC data, see Table \ref{Table3}.

\begin{table}
 \caption{The heliocentric minima times as derived from the INTEGRAL data.}
 \label{Table3} \centering \scalebox{0.80}{
\begin{tabular}{ c c c c }
\hline \hline
          &   HJD      & Error & Type \\[-2mm]
   Star   &  2400000+  & [days]&      \\ \hline
 V479 Aql & 53092.199  & 0.003 & prim \\[-2mm]
 V479 Aql & 53308.096  & 0.002 & prim \\[-2mm]
 V646 Cas & 53552.4654 & 0.004 & prim \\[-2mm]
 V646 Cas & 53564.7782 & 0.003 & prim \\[-2mm]
 V646 Cas & 53919.1205 & 0.01  & sec \\[-2mm]
 V646 Cas & 54547.7201 & 0.011 & sec \\[-2mm]
 V646 Cas & 54140.9981 & 0.004 & sec \\[-2mm]
 PV Cen   & 54118.589  & 0.003 & prim \\[-2mm]
 EP Cen   & 54642.0452 & 0.001 & prim \\[-2mm]
 EP Cen   & 54659.6262 & 0.001 & prim \\[-2mm]
 EP Cen   & 54670.7307 & 0.002 & prim \\[-2mm]
 RW Cet   & 53925.9339 & 0.0005& prim \\[-2mm]
 RW Cet   & 54640.7550 & 0.0005& prim \\[-2mm]
 RR Cir   & 52834.3033 & 0.001 & prim \\[-2mm]
 RR Cir   & 52837.0403 & 0.004 & sec \\[-2mm]
 RR Cir   & 54642.1967 & 0.003 & prim \\[-2mm]
 UX Cir   & 53372.7898 & 0.006 & prim \\[-2mm]
 BB Cir   & 53217.4341 & 0.003 & prim \\[-2mm]
 V445 Cyg & 53968.8264 & 0.002 & prim \\[-2mm]
 V445 Cyg & 54060.3704 & 0.002 & prim \\[-2mm]
 SX Gem   & 53111.5577 & 0.0002& prim \\[-2mm]
 SX Gem   & 53113.6186 & 0.002 & sec \\[-2mm]
 AB Mus   & 54683.9468 & 0.004 & prim \\[-2mm]
 AB Mus   & 53376.4548 & 0.002 & prim \\[-2mm]
 FL Ori   & 54327.6181 & 0.001 & prim \\[-2mm]
 SW Pup   & 52980.5239 & 0.009 & sec \\[-2mm]
 SW Pup   & 53858.3660 & 0.01  & prim \\[-2mm]
 V501 Sgr & 54023.6718 & 0.006 & prim \\[-2mm]
 V501 Sgr & 54034.2598 & 0.005 & prim \\[-2mm]
 V501 Sgr & 54046.3703 & 0.006 & prim \\[-2mm]
 V501 Sgr & 54050.9075 & 0.004 & prim \\[-2mm]
 EL Vel   & 52979.0020 & 0.009 & sec \\
 \hline
\end{tabular}}
\end{table}

\section{Discussion and conclusions}

The light-curve analyses of twenty-one selected systems have been carried out. The light curves
observed by the Optical Monitoring Camera onboard the INTEGRAL satellite provide a great
opportunity to study and to estimate the basic physical parameters of these systems. Despite this
fact, the parameters are still only the preliminary ones, affected by relatively large errors and
some of the relevant parameters were fixed at their suggested values. The detailed analysis is
still needed, especially spectroscopic one, or another more detailed light curve study in different
filters. Together with a prospective radial-velocity study, the final picture of these systems
could be done. Particularly, the systems PV~Cen and V501~Sgr seem to be the most interesting ones
due to relatively high value of the third light. Moreover, the system V646~Cas seems to be a very
massive binary, while on the other hand the system V2383~Oph is a low-mass binary of K spectral
type.

\section{Acknowledgments}
Based on data from the OMC Archive at LAEFF, pre-processed by ISDC. This investigation was
supported by the Czech Science Foundation grant no. P209/10/0715 and also by the Research Program
MSM 0021620860 of the Ministry of Education of Czech Republic and also by the Mexican grant PAPIIT
IN113308. This research has made use of the SIMBAD database, operated at CDS, Strasbourg, France,
and of NASA's Astrophysics Data System Bibliographic Services.

\end{document}